\documentclass[twoside]%,10pt]%[12pt]
{article}
\usepackage{myqic,amssymb,euscript,verbatim}
\newcommand{\enproof}{\hfill $\Box$ \vspace*{1ex}}
\newcommand{\enlem}{\hfill $\Diamond$ \vspace*{1ex}\end{lemma}}
\newcommand{\closedef}{\hfill $\Diamond$ \end{definition}}
\newcommand{\enth}{\hfill $\Diamond$ \end{theorem}}
\newcommand{\encor}{\hfill $\Diamond$ \end{corollary}}
\newcommand{\enprop}{\hfill $\Diamond$ \end{proposition}}
\newcommand{\encond}{\hfill $\Diamond$ \end{condition}}
\newcommand{\exam}[1]{\begin{example}\label{ex:#1}}
\newcommand{\enexam}{\QED\end{example}}
\newcommand{\beremark}[1]{\begin{remark}\label{rmk:#1}}

\newcommand{\mymathbb}[1]{{\mathbb #1}} 
\newcommand{\mymathsf}[1]{{\mathsf{#1}}} 
\newcommand{\mymathcal}[1]{{\EuScript #1}} %from '02 Dec %IEICE myPC
\renewcommand{\cal}{\EuScript} %from '02 Dec.

\newcommand{\cA}{{\cal A}}

\newcommand{\cB}{{\cal B}}

\newcommand{\cC}{{\cal C}}

\newcommand{\cD}{{\cal D}}

\newcommand{\sH}{\mymathcal{H}} %{\mymathsf{H}}  %'02 Dec.
\newcommand{\sK}{\mymathcal{K}} %{\mymathsf{K}}  %'02 Dec.
\newcommand{\cI}{{\cal I}}

\newcommand{\cM}{{\cal M}}
\newcommand{\sM}{\mymathsf{M}}

\newcommand{\cR}{{\cal R}}

\newcommand{\cT}{{\cal T}}

\newcommand{\cX}{{\cal X}}
\newcommand{\sX}{\mymathsf{X}}
\newcommand{\cY}{{\cal Y}}

\newcommand{\men}[1]{\Phi_{#1}}

\newcommand{\vep}{\varepsilon}
\renewcommand{\phi}{\varphi}
\renewcommand{\subset}{\subseteq}

\renewcommand{\hat}{\widehat}

\newcommand{\SINT}{\mymathbb{Z}}

\newcommand{\Capa}{\mymathsf{Q}} %{\mymathsf{C}} 
\newcommand{\Expo}{\mymathsf{E}}
\newcommand{\Dopt}{\mymathsf{D}} %{\mymathsf{C}} 

\newcommand{\Prob}{{\rm Pr}}

\newcommand{\tnsr}{\otimes}
\newcommand{\trace}{{\rm Tr}}

\newcommand{\lag}{\langle}
\newcommand{\rag}{\rangle}
\newcommand{\crd}[1]{|#1|}
\newcommand{\bra}[1]{\lag #1 |}
\newcommand{\ket}[1]{| #1 \rag}

\newcommand{\syp}[2]{( #1,  #2 )_{\rm sp}}

\newcommand{\dmn}{d}
\newcommand{\Hch}{{\sH}}
\newcommand{\Ha}{\sH_{\rm A}} %i A
\newcommand{\Hb}{\sH_{\rm B}} %o B
\newcommand{\Ht}{\sH_{\rm T}} %o B
\newcommand{\Hgn}{{\sH}}
\newcommand{\Hgen}{{\sH}}
\newcommand{\Hdis}[1]{\sH_{#1}}

\newcommand{\Han}{\Ha^{\tnsr n}}
\newcommand{\Hbn}{\Hb^{\tnsr n}}
\newcommand{\Htn}{\Ht^{\tnsr n}}
\newcommand{\Hchn}{\Hch^{\tnsr n}}
\newcommand{\Bop}{\mymathcal{L}} %{\sL} %'02 Dec.
\newcommand{\Hcd}{\cC}
\newcommand{\rC}{C}
\newcommand{\Cq}{\Hcd}

\newcommand{\Fen}{F_{\rm e}}
\newcommand{\Ebe}{N}
\newcommand{\ketbe}[1]{\ket{#1}}
\newcommand{\phasebe}{\omega}
\newcommand{\Xbe}{X}
\newcommand{\Zbe}{Z}

\newcommand{\hEnk}[3]{\hat{E}_{#3}} %_{#1,#2}}
 
\newcommand{\CPex}{\cM}
\newcommand{\CPexO}{M}
\newcommand{\Cso}{L}

\newcommand{\varnin}{n} %4 inner codes %4switch used to be \nu
\newcommand{\varkin}{k} %4 inner codes %4switch used to be \kappa
\newcommand{\spn}{\mymathsf{span}\,}
\newcommand{\hPi}{\hat{\Pi}}
\newcommand{\Id}{\cI} %{{\rm I}}
\newcommand{\varM}{n} %{m} %'02 Dec.

\newcommand{\TlK}{T}
\newcommand{\Tel}{\cT}
\newcommand{\Dis}{\cD}
\newcommand{\dec}{\cR} %{\cD}%
\newcommand{\tch}[1]{\chi_{#1}} 
\newcommand{\hM}[1]{\sM({#1})} 
\newcommand{\shs}{\rho} %shared state used be sigma
\newcommand{\DOgen}{\sigma} %general density operator
\newcommand{\Fdis}[2]{F^{\star}(#2,#1)} %(#2;#1)} %_{\rm dis}
\newcommand{\Fdiscl}[3]{F_{#1}^{\star}(#3,#2)} %(#3;#2)} %_{\rm dis}
\newcommand{\Fch}[3]{F_{\rm ch}^{\star}(#3,#2)} %(#3;#2)}
\newcommand{\Zdpower}[2]{\SINT_{#1}^{#2}}

\begin{document}
\setlength{\textheight}{8.0truein}    %FOR 2ND PAGE ONWARDS

\runninghead{Teleportation and entanglement distillation
for correlated mixed states}{M.~Hamada} % $\ldots$}

\normalsize\textlineskip
\thispagestyle{empty}
\setcounter{page}{1}

%\copyrightheading{Vol.}{No.}{Year}{Page Nos.}
%\copyrightheading{0}{0}{2003}{000--000}

%\vspace*{0.88truein}

\alphfootnote

\fpage{1}

\centerline{\bf
%%%%%%%%%%%%%%%%%%%%%
%Put in titiles here
%%%%%%%%%%%%%%%%%%%%%
TELEPORTATION AND ENTANGLEMENT DISTILLATION}
\vspace*{0.035truein}
\centerline{\bf 
IN THE PRESENCE OF CORRELATION} 
\vspace*{0.035truein}
\centerline{\bf 
AMONG BIPARTITE MIXED STATES}
%Teleportation and entanglement distillation
%in the presence of
%correlation among bipartite mixed states
%\footnote{Typeset the title in 10 pt Times Roman, uppercase and boldface.}}
\vspace*{0.37truein}
\centerline{\footnotesize
%%%%%%%%%%%%%%%%%%%%%%%%%%%%%%%%%%%%
%put authors' name and address here
%%%%%%%%%%%%%%%%%%%%%%%%%%%%%%%%%%%%
MITSURU HAMADA}
%\footnote{Typeset names in
%10 pt Times Roman, uppercase. Use the footnote to indicate the
%present or permanent address of the author.}
%\vspace*{0.015truein}
\vspace*{0.020truein}
\baselineskip=10pt
\centerline{\footnotesize\it
Quantum Computation and Information Project (ERATO)}
\vspace*{0.015truein}
\centerline{\footnotesize\it
     Japan Science and Technology Corporation}
%\eads{mitsuru@ieee.org}
\vspace*{0.015truein}
\centerline{\footnotesize\it 
              %201 Daini Hongo White Bldg.,\
      5-28-3, Hongo, Bunkyo-ku, Tokyo 113-0033, Japan}
\vspace*{10pt}
\vspace*{0.225truein}
%\publisher{(received date)}{(revised date)}

%\vspace*{0.21truein}

%% \abstracts{first paragraph}{second paragraph}{third paragraph}
%% If there is only one paragraph, just keep the second and third empty 
%% like the following one 
\abstracts{
%%%%%%%%%%%%%%%%%%%%
% put abstract here
%%%%%%%%%%%%%%%%%%%%
%\begin{abstract}
The teleportation channel
associated with an arbitrary bipartite state
denotes the map that represents the change suffered by a teleported state
when the bipartite state is used instead of the ideal
maximally entangled state for teleportation.
This work presents and proves
an explicit expression of the teleportation channel
for the teleportation using Weyl's projective unitary representation
of $(\SINT/\dmn\SINT)^{2n}$ for integers $\dmn \ge 2,n \ge 1$,
which has been known for $n=1$.
This formula allows any correlation among the $n$ bipartite
mixed states, 
%shared by two parties, 
and an application shows the existence of reliable schemes for
distillation of entanglement from a sequence of mixed states with correlation.
\iffalse
For example,  distillation with a positive asymptotic rate
is shown to be possible of
a sequence of Bell states $\ket{00}\pm\ket{11}, \ket{01}\pm\ket{10}$
which occur according to the probability measure of a Markov chain.
\fi
%\end{abstract}
}{}{}

\vspace*{10pt}

%\vspace*{3pt}
%\communicate{to be filled by the Editorial}

\vspace*{1pt}\textlineskip    %) USE THIS MEASUREMENT WHEN THERE IS
   %) A SECTION HEADING
%\vspace*{-0.5pt}
%\noindent
%%%%%%%%%%%%%%%%%%%%%%%%%%%%%%%%
%put the text of the paper here
%%%%%%%%%%%%%%%%%%%%%%%%%%%%%%%%

\begin{comment}
\address{Quantum Computation and Information Project (ERATO)\\
     Japan Science and Technology Corporation\\
              %201 Daini Hongo White Bldg.,\
      5-28-3, Hongo, Bunkyo-ku, Tokyo 113-0033, Japan}
\eads{mitsuru@ieee.org}
\end{comment}

%\date{Jan., 2002}

%\begin{document} %iop

%\maketitle %iop

\iffalse %iop
\fi

\section{Introduction \label{ss:intro}}

Relationships between
entanglement distillation and quantum error correction
have been discussed by Bennett {\em et al.}\/~\cite{bennett96m}.
Especially, they argued that achievable information rates for 
quantum error correction, i.e., those 
at which quantum error-correcting codes
(quantum codes) reliably, 
are also achievable as rates for one-way entanglement distillation.
More precisely, they associate with
an arbitrary bipartite mixed state a map called a teleportation channel,
which represents the change suffered by a teleported state 
when the bipartite mixed state
is used for teleportation~\cite{bennett93}
in place of the ideal maximally entangled state.
Then, they argued that an achievable rate for quantum codes on
the teleportation channel
is also achievable as the asymptotic yield of distillation schemes for the
bipartite state.
A concrete expression for the teleportation channel using $(\SINT/\dmn\SINT)^2$
was given afterwards~\cite{BowenBose01}, which complements the above
argument on transformation of achievable rates. 
Recently, 
this author gave exponential lower bounds on the highest fidelity
and those on the largest information rates
that can be attained by standard algebraic %4SPACE
quantum codes~\cite{Gottesman96,crss97}
%, from which lower bounds on the quantum capacity immediately follows,
not only on discrete memoryless quantum channels
but also on channels with certain correlation~\cite{hamada01e,hamada01g,hamada02m,hamada02c}.
This work was motivated by interest in exploring implications
of these results~\cite{hamada01e,hamada01g,hamada02m,hamada02c}
on entanglement distillation, especially in the presence of correlation
among the bipartite %mixed 
states,
along the lines of \cite{bennett96m}. %4SPACE_CANDIDATE

In what follows, we will do the next three things. 
(i) %({\em a}\/) 
To deal with correlated states,
the formula for the teleportation channel using 
$(\SINT/\dmn\SINT)^{2}$~\cite{BowenBose01} is generalized
to that for teleportation using Weyl's projective unitary representation
of $(\SINT/\dmn\SINT)^{2n}$,
which allows any correlation among the $n$ mixed states shared by two parties,
and proved in such a way that the role of (characters of) the underlying group
$(\SINT/\dmn\SINT)^{2n}$ becomes clear.
(ii) %({\em b}\/) 
We refine Bennett {\em et al.}\/'s observation~\cite{bennett96m}.
Namely, while they have discussed only asymptotically achievable rates,
we will directly work with fidelity,
and show that
trade-offs between the fidelity and rates of quantum codes
can be transformed
into those between the fidelity and rates of one-way distillation protocols.
(iii) %({\em c}\/) 
We apply these arguments to the known %achievability 
results on quantum codes~\cite{hamada01e,hamada01g,hamada02m,hamada02c}.
Namely, we present exponential lower bounds on the largest fidelity
that can be attained by one-way distillation 
protocols using %by means of 
the generalized formula in (i) and transformations in (ii).
For example,  reliable distillation with a positive asymptotic rate
and exponential decay of unity minus fidelity
is shown to be possible of
a sequence of Bell states $\ket{00}\pm\ket{11}, \ket{01}\pm\ket{10}$
which occur according to the probability measure of a Markov chain.

\section{Notation and Basic Notions \label{ss:pre}}

%First, we fix notation and basic notions on entanglement distillation
%and quantum error correction. %channel coding.

\subsection{Entanglement Distillation}

The set of all linear operators on a Hilbert space $\Hgen$ is denoted
by $\Bop(\Hgen)$. As usual, 
states are represented as unit-trace positive semidefinite element of
$\Bop(\Hgen)$ or sometimes, when they are pure, normal vectors in $\Hgen$.
For a trace-preserving completely positive (TPCP) 
linear map $\CPex: \Bop(\Hgn) \to \Bop(\Hgn)$, 
we write $\CPex \sim \{ \CPexO_i \}_{i\in\cY}$ if
$\CPex(\DOgen) = \sum_{i\in\cY} \CPexO_i \DOgen \CPexO_i^{\dagger}$
(an operator-sum representation).
The identity in $\Bop(\Hgen)$ is denoted by $I$ while
the identity map from $\Bop(\Hgen)$ onto $\Bop(\Hgen)$ is denoted by $\Id$.
We distinguish 
$I$ (or $\Id$) for different Hilbert spaces, say $\Ha$,
$\Hgen_{n,{\rm R}}$, $\Htn\tnsr\Han$, {\em etc.,}\/
(by a rather loose notation) using subscripts, say, A, R, TA, {\em etc.,}\/
as in $I_{\rm TA}$.

Let $\Ha$ and $\Hb$ be Hilbert spaces of finite dimensions.
Without loss of generality, we assume $\dim\Ha=\dim\Hb$ ($=\dmn$).
Our purpose is to distill a bipartite mixed state $\shs_n$ 
in $\Bop(\Ha^{\tnsr n}\tnsr\Hb^{\tnsr n})$
into a maximally entangled state 
\begin{equation}\label{eq:me}
\ket{\men{\cB_{\rm A},\cB_{\rm B}}} = \frac{1}{K} \sum_{0\le j<K}\ket{j}_{\rm
A}\tnsr\ket{j}_{\rm B},
\end{equation}
with some Hilbert spaces $\Hdis{n, {\rm A}},\Hdis{n, {\rm B}}$,
and orthonormal systems 
$\cB_{\rm A} = \{ \ket{j}_{\rm A} \} \subset \Hdis{n, {\rm A}},
%$ and $
\cB_{\rm B} = \{ \ket{j}_{\rm B} \} \subset \Hdis{n, {\rm B}}$
by a TPCP linear map
\[
\Dis_n:\, \Bop(\Ha^{\tnsr n}\tnsr\Hb^{\tnsr n}) \to \Bop(\Hdis{n,{\rm
A}} \tnsr \Hdis{n,{\rm B}}),
\]
where $\Dis_n$ is in some presupposed class $C$ of distillation protocols.
For any such pair $(\cB_{\rm A}, \cB_{\rm B})$,
we say $\ket{\men{\cB_{\rm A},\cB_{\rm B}}}$ is distillable or
$C$-distillable from $\shs_n$ with fidelity 
\[
\bra{\men{\cB_{\rm A},\cB_{\rm B}}} \Dis_n( \shs_n ) \ket{\men{\cB_{\rm A},\cB_{\rm B}}}.
\]
We want both fidelity
and $\crd{\cB_{\rm A}}=\crd{\cB_{\rm B}}$ to be
as large as possible, but there is a trade-off between them,
which have been investigated
from an information theoretic interest~\cite{rains98,rains01}. 
Thus, we will estimate
\begin{eqnarray}
\Fdis{K}{\shs_n}&=&\Fdiscl{C}{K}{\shs_n} \nonumber\\
&=& \sup \left\{ F \,\left|
\begin{array}{l}
\mbox{a state $\ket{\men{\cB_{\rm A},\cB_{\rm B}}}$ with $|\cB_{\rm
A}|=|\cB_{\rm B}| \ge K$ is}\\
\mbox{$C$-distillable from $\shs_n$ with fidelity $F$}
\end{array}\right.
\right\}
\end{eqnarray}
for a state $\shs_n\in \Bop(\Han\tnsr\Hbn)$,
where $0< K \le \dmn^n$ is a real number.

Our protocol using quantum codes to be discussed below
is a simple one-way distillation protocol,
which consists of one measurement on the (possibly locally
enlarged) site A,
and a local quantum operation (TPCP linear map) on the site B
that may be chosen according to the measurement result on the site A.
We may suppose $C$ denotes the class of such protocols,
but we assume $C$ is the slightly more general class
$\rC_1$ of 1-local operations
following Rains' lucid classification~\cite{rains01}.
Namely, unless otherwise mentioned, $C=\rC_1$ is assumed throughout and
the subscript $C$ in $\Fdiscl{C}{K}{\shs_n}$ or similar quantities
will be suppressed.

It is remarked that requiring $\Hdis{n,{\rm A}}=\spn \cB_{\rm A}$ and
$\Hdis{n,{\rm B}}=\spn \cB_{\rm B}$ in defining `$C$-distillable',
as in \cite{rains01}, does not change the above quantity $\Fdis{K}{\shs_n}$
when local measurements are allowed as in the present case where $C=\rC_1$.
%
\begin{comment}
For, do a measurement ${\Pi_B, I-\Pi_B}$ where $B=\Hdis{n,{\rm B}}$ on site B.
\end{comment}
%
We will not enter into details of classification of distillation
protocols, but only mention that $C=\rC_1$ represents the class of what
we call one-way distillation protocols, $C$ is contained in
(but significantly differs in capability~\cite{bennett96p,bennett96m} from)
the class of two-way protocols, they are separable
and hence positive partial transpose (PPT) ones~\cite{rains98,rains01}.

\subsection{Quantum Error Correction}

%{\em Channel coding.}\/

Hereafter throughout the paper, 
$\Hch$ is a Hilbert space with dimension $\dmn\ge 2$.
We will consider the situation where quantum states are transmitted
through a quantum channel with input-output Hilbert space $\Hch$,
which is a sequence of TPCP
linear maps $\{ \cA_n \}$ with
\[
\cA_n: \, \Bop(\Hch^{\tnsr n}) \to \Bop(\Hch^{\tnsr n}).
\]
The map $\cA_n$ with fixed $n$ is sometimes called a channel also.
A quantum code is a pair $(\Cq_n, \dec_n)$
that consists of a subspace $\Cq_n \subset \Hch^{\tnsr n}$ and a decoder,
which is a TPCP linear map,
\begin{eqnarray*}
\dec_n:\, \Bop(\Hch^{\tnsr n}) \to \Bop(\Hch^{\tnsr n}).
\end{eqnarray*}
The subspace $\Cq_n$ alone is also called a quantum code.%
\footnote{Whereas we treat this class of coding schemes for simplicity,
Theorem~\ref{th:main} below holds true if a more general class
(those with encoding maps) are allowed as in \cite{bennett96m,barnum00},
which will be clear from the proof.} %QIC
\mbox{ }While there are two fidelity measures often used 
for evaluating quantum codes, i.e., {\em entanglement fidelity}\/ and
{\em minimum pure-state fidelity}\/ (minimum fidelity),
we will mostly work with the entanglement fidelity
since, as its name and definition suggest,
it is directly related to entanglement distillation. 
\begin{comment}
Entanglement fidelity measures the amount of entanglement that is shared between two parties,
say, R and B,
after B receives a half of an entangled state
through noisy channels while R's part is undisturbed~\cite{schumacher96}.
More precisely, 
\end{comment}
Entanglement fidelity is defined as follows~\cite{schumacher96}.
Let $\Pi_{\Cq}$ and $\hPi_{\Cq}$ denote
the projection onto $\Cq$ and its normalization
$
(\dim \Cq)^{-1} \Pi_{\Cq},
$
respectively.
Then, given a quantum code as above with $\dim \Cq_n =K$, 
we prepare a $K$-dimensional Hilbert space $\Hch_{n, {\rm R}}$,
and a maximally entangled state
\[
\ket{\men{\cB_{\rm R},\cB}}= \frac{1}{K} \sum_{0\le j<K}\ket{j}_{\rm
R}\tnsr\ket{j},
\]
where $\cB_{\rm R}=\{ \ket{j}_{\rm R} \}_{0\le j<K}
\subset \Hch_{n,{\rm R}}$ and
$\cB = \{ \ket{j} \}_{0\le j< K} \subset \Hch^{\tnsr n}$ are systems
of orthonormal vectors, and define the entanglement fidelity 
$\Fen( \hPi_{\Cq}, \cM )$ for the state $\hPi_{\Cq}$
and a general TPCP map $\cM$ by
\[
\Fen( \hPi_{\Cq}, \cM  ) =  
\bra{\men{\cB_{\rm R},\cB}}[\Id_{\rm R} \tnsr \cM](\ket{\men{\cB_{\rm
R},\cB}}\bra{\men{\cB_{\rm R},\cB}}) \ket{\men{\cB_{\rm R},\cB}}.
\]
The entanglement fidelity $\Fen( \hPi_{\Cq}, \cM )$
does not depend on the choice of the purification $\ket{\men{\cB_{\rm
R},\cB}}$ of $\hPi_{\Cq}$~\cite{schumacher96}.
Trade-offs between $\Fen( \hPi_{\Cq}, \dec\circ\cA_n  )$ and $\dim \Cq$
have been investigated in the literature.
Namely, the quantity of interest has been
\begin{equation}
\Fch{n}{K}{\cA_n}
= \sup \left\{ F \,\left|
\begin{array}{l}
\mbox{a quantum code $(\Cq,\dec)$ exists with}\\
\mbox{$\Fen( \hPi_{\Cq}, \dec\circ\cA_n ) = F$ and $\dim \Cq \ge K$}
\end{array}\right. \right\}.
\end{equation}

\subsection{Capacity, %Error Exponents 
Reliability Function and Their Analogues for Distillation}

Given a channel $\{ \cA_n \}$, 
%a nonnegative %4SPACE
a number $R \ge 0$ is said to be an achievable rate on $\{ \cA_n \}$
if it satisfies
\[
\lim_{n\to\infty} \Fch{n}{\dmn^{Rn}}{\cA_n}=1.
\]
The supremum of achievable rates on $\{ \cA_n \}$
is called the {\em quantum capacity}\/ of $\{ \cA_n \}$ and
denoted by $\Capa(\{ \cA_n \})$.
A function $E(R) \ge 0$ is said to be an attainable (or achievable)
error exponent if it satisfies
\[
1-\Fch{n}{\dmn^{Rn}}{\cA_n} \le \exp_{\dmn} [-n E(R) +o(n) ]
\]
and  the pointwise supremum of attainable exponents on $\{ \cA_n \}$
is called the {\em reliability function}\/ (the optimum error exponent) of
$\{ \cA_n \}$ and denoted by $\Expo(R,\{ \cA_n \})$.

Turning to entanglement distillation, 
for a sequence of states $\{ \shs_n \}$, we can define
the counterparts of $\Capa( \{ \cA_n \})$ and $\Expo(R, \{ \cA_n \})$,
which this paper is concerned with. 
For example, the capacity analogue
$\Dopt(\{ \shs_n \})$ %=\Dopt_{C}(\{ \shs_n \})$
is defined as the supremum
of $R$ such that $\Fdis{\dmn^{nR}}{\shs_n}=1$.
In the literature, $\Dopt(\{ \shs^{\tnsr n} \})$
is sometimes called the distillable entanglement of $\shs$.
%
%Roughly speaking,
%for our choice of $C$, $\Dopt(\{ \shs^{\tnsr n} \})$ is the same as 
%$D_1( \shs )$ in \cite{bennett96m}.?
%

\subsection{Algebraic Structure Underlying Teleportation and Symplectic Codes \label{ss:Weyl}}

Algebraic structures underlie quantum mechanical phenomena~\cite{weyl31}
as well as schemes for quantum information processing
such as teleportation protocols and symplectic (stabilizer) codes~\cite{Gottesman96,crss97}. %,crss98},
In such quantum operations, the structure of
$\SINT_\dmn=\SINT/\dmn\SINT$ is exploited,
and its link to familiar Hilbert spaces is given 
by a projective (ray) representation $\Ebe:\, u\mapsto\Ebe_u$ 
of $\SINT_\dmn \times \SINT_\dmn$,
which is defined as follows~\cite{weyl31}.
Fix an orthonormal basis 
$\{ \ketbe{0},\dots, \ketbe{\dmn-1} \}$ of $\Hch$,
put $\cX=\SINT_\dmn\times\SINT_\dmn$,
and 
\[
\Ebe_{(i,j)} = \Xbe^i \Zbe^j, \quad (i,j)\in \cX,
\]
where 
$\Xbe, \Zbe \in \Bop(\Hch)$ are defined by
\begin{equation}\label{eq:error_basis}
\Xbe \ketbe{j}  = \ketbe{j-1}, \,\,\,
\Zbe \ketbe{j} = \phasebe^ j \ketbe{j}, \quad \,\,\, j\in\SINT_\dmn
\end{equation}
with $\phasebe$ being a primitive $\dmn$-th root of unity.
When $\dmn=2$, this representation is essentially the same as
the system of Pauli matrices with identity.

To deal with the larger system of $\Hch^{\tnsr n}$, we put
\begin{equation}\label{eq:XZomega}
\Ebe_y = \Ebe_{y_1} \tnsr \dots \tnsr \Ebe_{y_n}
\end{equation}
for $y=(y_1,\dots,y_n)\in \cX^n$.
We identify $\cX^n$ with $\Zdpower{\dmn}{2n}$
via the trivial correspondence
\[
((x_1,z_1),\dots, (x_n,z_n)) \mapsto
(x_1,z_1,\dots,x_n,z_n).
\]
Observe the commutation relation
\begin{equation}\label{eq:commutative}
 \Ebe_{y} \Ebe_{y'} = \omega^{\syp{y}{y'}} \Ebe_{y'} \Ebe_{y},
\end{equation}
where
\begin{equation}\label{eq:syp}
\syp{y}{y'} = \sum_{i=1}^{n} x_i z_i' - z_i x_i' 
\end{equation}
for $y=(x_1,z_1,\dots,x_n,z_n)$ and 
$y'=(x'_1,z'_1,\dots,x'_n,z'_n) \in \Zdpower{\dmn}{2n}$.

\section{Teleportation Channel \label{ss:telechan}}

It is known that
for every unitary basis of $\Bop(\Hgen)$ which is orthonormal with respect to
the normalized Hilbert-Schmidt inner product,
there exists a teleportation protocol using it~\cite{werner01}.
Since Weyl's basis
$\{ \Ebe_y \}_{y\in\cX^n}$ is such a unitary basis~\cite{schwinger60},
we have a teleportation protocol using it that teleports states in $\Hchn$.
Note that 
\[
\ket{\Psi_{y}} = 
\frac{1}{\dmn^{n/2}} \sum_{l\in\Zdpower{\dmn}{n}} \ket{l} \tnsr \Ebe_y \ket{l}, \quad y \in \Zdpower{\dmn}{2n},
\]
and
\[
\ket{\Psi'_x}= \frac{1}{\dmn^{n/2}}\sum_{l\in\Zdpower{\dmn}{n}} \Ebe_x \ket{l} \tnsr \ket{l}, \quad x \in \Zdpower{\dmn}{2n},
\]
where $\ket{(l_1,\dots,l_n)}=\ket{l_1}\tnsr\dots\tnsr\ket{l_n}$, 
form orthonormal bases of $\Hgen^{\tnsr n} \tnsr \Hgen^{\tnsr n}$~\cite{werner01}.

We will present a concrete expression of the teleportation channel for
Weyl's basis $\{ \Ebe_y \}_{y\in\cX^n}$, 
which allows us to treat correlated states.
We will prove this in a simple manner using a property of the underlying
additive group $\Zdpower{\dmn}{2n}$ for arbitrary integers
$\dmn \ge 2, n \ge 1$.

\begin{figure}
\begin{center}
\centerline{\epsfig{{file=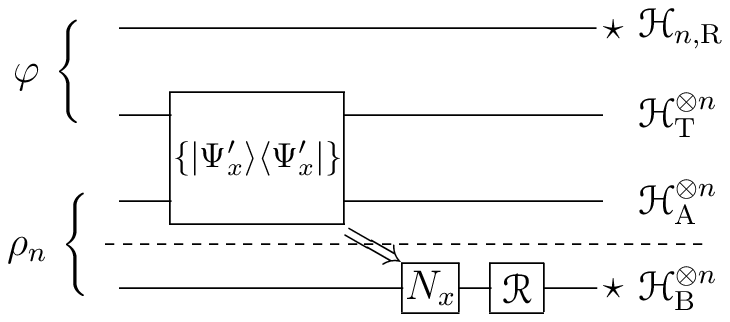},width=6.4cm}}
\mbox{}
\end{center}
\fcaption{Entanglement distillation using a quantum code $(\Hcd,\dec)$.
%qic
The state $\phi$ is a purification of $\DOgen=\hPi_{\Hcd}$
to be teleported, 
and $\shs_n$ is the shared bipartite state for quantum teleportation.
The unitary $\Ebe_x$ is chosen according to the result of the measurement 
$\{ \ket{\Psi'_x}\bra{\Psi'_x} \}_{x\in\SINT_\dmn^{2n}}$
on site A, which is
sent to B by classical communication.
After the recovery operation $\dec$ of the code, an entangled 
state is shared between the positions indicated by $\star$.
 \label{fig:1}}
%\end{center}
\end{figure}

To describe the protocol,
we prepare $\Ha^{\tnsr n}$, $\Hb^{\tnsr n}$ and $\Ht^{\tnsr n}$,
where the Hilbert spaces $\Ha$, $\Hb$ and $\Ht$ have the same
dimension $\dmn$ 
(see Fig.~\ref{fig:1} ignoring $\Hgen_{n,{\rm R}}$ and $\dec$).
We will often identify $\Ha$, $\Hb$ and $\Ht$ with $\Hgen$ having a basis
$\{ \ket{j} \}_{j\in\SINT_\dmn}$.
Given a bipartite state $\shs_n$ in $\Bop(\Han\tnsr\Hbn)$,
sender A teleports a state in $\Ht^{\tnsr n}$ to receiver B.
The entire process of the teleportation on the systems 
of A, B and T can be represented by
an operator-sum representation
\begin{equation}\label{eq:tel0}
\Tel(\DOgen\tnsr\shs_n) = 
\sum_{x\in\Zdpower{\dmn}{2n}} \TlK_{x} (\DOgen \tnsr \shs_n) \TlK_{x}^{\dagger}
\end{equation}
where $\DOgen\in\Bop(\Htn)$ is the state to be teleported, and
\begin{equation}\label{eq:tel1}
\TlK_{x} = (I_{\rm TA} \tnsr \Ebe_{x}) 
(\ket{\Psi'_{x}} \bra{\Psi'_{x}}
\tnsr I_{\rm B} )
= \ket{\Psi'_{x}} \bra{\Psi'_{x}} \tnsr \Ebe_{x}.
\end{equation}
\iffalse
%This teleportation is designed so that it becomes perfect for
%$\shs_n=(1/\dmn^n) \sum_{l\in
%\Zdpower{\dmn}{n}} \ket{l}\tnsr\ket{l}$ as we will see soon.
\fi

It is natural to ask 
how $\DOgen$ changes during the teleportation
for arbitrary bipartite mixed states $\shs_n$.
This state change, which is of course a TPCP linear map,
will be denoted by $\tch{\shs_n}$ and called a teleportation channel.
The next lemma answers the question.
\begin{lemma}\label{lem:telechan}
The teleportation channel $\tch{\shs_n}$ for the above protocol is given by
\[
\tch{\shs_n}(\DOgen) = \sum_{x \in\Zdpower{\dmn}{2n}}
\bra{\Psi_{x}} \shs_n \ket{\Psi_{x}} 
\Ebe_x \DOgen \Ebe_x^{\dagger} .
\]
\end{lemma}

{\em Proof.}\/
It is enough to calculate (\ref{eq:tel0}) for $\DOgen=\ket{u}\bra{v}$,
$u,v\in\Zdpower{\dmn}{n}$.
Write $\shs_n$ as
\begin{equation}\label{eq:MatrixBellBasis}
\shs_n = \sum_{y,z\in\Zdpower{\dmn}{2n}} \alpha_{y,z} 
\ket{\Psi_{y}} \bra{\Psi_{z}},
\end{equation}
and note that by (\ref{eq:error_basis}),
\begin{equation}\label{eq:nEbe}
\Ebe_{x} \ket{l} = \omega^{b\cdot l} \ket{l-a}
\end{equation}
for $x=(a_1,b_1,\dots,a_n,b_n)$,
$a=(a_1,\dots,a_n)$, $b=(b_1,\dots,b_n)$,
$l=(l_1,\dots,l_n)$, and 
\[
b\cdot l=\sum_{i}b_i l_i.
\]
Then, (\ref{eq:tel1}) can be rewritten as 
\begin{equation}
\TlK_{x} = \frac{1}{\dmn^{n}}
\sum_{l,m} \omega^{b\cdot (l-m)}
\ket{l-a}\bra{m-a} \tnsr \ket{l}\bra{m}\tnsr \Ebe_x,
\end{equation}
with which the summand in (\ref{eq:tel0})
can be calculated as
\begin{eqnarray*}
\TlK_{x} (\ket{u}\bra{v} \tnsr \shs_n) \TlK_{x}^{\dagger}&=&
\frac{1}{\dmn^{3n}}
\sum_{l,l'} \omega^{b\cdot (l-u-l'+v)} \ket{l-a}\bra{l'-a}\\
&&\mbox{}\tnsr \ket{l}\bra{l'} 
\tnsr \sum_{y,z} \alpha_{y,z} 
\Ebe_{x}\Ebe_{y} \ket{u+a}\bra{v+a} \Ebe_z^{\dagger}\Ebe_x^{\dagger}.
\end{eqnarray*}
Using (\ref{eq:commutative}) and (\ref{eq:nEbe}), we then have
\begin{equation}\label{eq:Tx}
\TlK_{x} (\ket{u}\bra{v} \tnsr \shs_n) \TlK_{x}^{\dagger}
= \frac{1}{\dmn^{2n}}
\Big(
\ket{\Psi'_x}\bra{\Psi'_x}
\tnsr \sum_{y,z} \alpha_{y,z} \omega^{\syp{x}{y-z}} \Ebe_{y}\ket{u}\bra{v}
\Ebe_z^{\dagger}
\Big).
\end{equation}
We take partial trace of the both sides of (\ref{eq:Tx})
over the systems of T and A, and
sum them over $x\in\Zdpower{\dmn}{2n}$
noting that 
\begin{equation}\label{eq:chara}
\sum_{x\in\Zdpower{\dmn}{2n}}
\omega^{\syp{x}{y-z}}=0 \quad \mbox{whenever} \quad y\ne z,
\end{equation}
which holds because $f_{y-z}: x\mapsto\omega^{\syp{x}{y-z}}$
is a {\em character}\/ of $\Zdpower{\dmn}{2n}$ (see the next paragraph).
Thus, we find the teleportation channel $\tch{\shs_n}$ is
\[
\tch{\shs_n}(\DOgen) = \sum_{y \in\Zdpower{\dmn}{2n}} \alpha_{y,y} \Ebe_y \DOgen 
\Ebe_y^{\dagger},
\]
as claimed. \enproof

We include a short proof of (\ref{eq:chara}),
though it is merely a property of characters of groups.
Put $f(x)=f_a(x)=\omega^{\syp{x}{a}}$, $a\in\Zdpower{\dmn}{2n}$. 
Then, for any $x'$,
\[
f(x') \sum_{x\in\Zdpower{\dmn}{2n}} f(x) =
\sum_{x\in\Zdpower{\dmn}{2n}} f(x'+x) = 
\sum_{x\in\Zdpower{\dmn}{2n}} f(x).
\]
Hence, if $f(x')\ne 1$ for some $x'\in\Zdpower{\dmn}{2n}$,
which is true for $a \ne 0$, we have
$\sum_{x\in\Zdpower{\dmn}{2n}} f(x)=0$, as desired.

It should be mentioned that there is an inverse process
that associate a bipartite mixed state with a channel (TPCP map)~\cite{bennett96m}. The map is
\begin{equation}\label{eq:choi}
\hM{\cA}= [\Id \tnsr \cA](\ket{\Psi_{0}} \bra{\Psi_{0}}).
\end{equation}
Its matrix is Choi's matrix divided by $\dmn^{n}$~\cite{choi75}.
Note that when $\shs_n$ is written in the form
(\ref{eq:MatrixBellBasis}), we have $\hM{\tch{\shs_n}}=\shs_n$
if and only if $(\alpha_{y,z})$ is diagonal.

It may be worth mentioning that another more straightforward
calculation using (\ref{eq:chara}) gives
a formula for a discrete `twirling',
which is effectively known~\cite[Appendix~A]{bennett96m}, \cite[Sec.~II.B]{VollbrechtW02}:
\begin{equation}\label{eq:twirling}
\frac{1}{\dmn^{2n}} \sum_{x\in\Zdpower{\dmn}{2n}}
(\overline{\Ebe_{x}} \tnsr \Ebe_x) \shs_n
(\overline{\Ebe_{x}} \tnsr \Ebe_x)^{\dagger}
=\sum_{y\in\Zdpower{\dmn}{2n}} \alpha_{y,y}\ket{\Psi_y}\bra{\Psi_y}
\end{equation}
for $\shs_n$ in (\ref{eq:MatrixBellBasis}), where
$\bra{l}\overline{\Ebe_{x}}\ket{m}$ is the complex conjugate of
$\bra{l}\Ebe_{x}\ket{m}$, $l,m\in\Zdpower{\dmn}{n}$.
In general, the resultant states are less disordered
(i.e., have less entropy)
than those obtained with the continuous or full twirling~\cite{bennett96m}.

\section{Using Quantum Codes for Entanglement Distillation \label{ss:QECC2D}}

Bennett {\em et al.}\/~\cite[p.3840, Sec.~V.C]{bennett96m} argued
%format [] OK?
that an achievable rate for a quantum channel by
quantum codes is also achievable as a distillation rate
for the corresponding bipartite states by one-way distillation protocols. 
In this section, we refine this argument
working directly with fidelity rather than achievable rates.

Bennett {\em et al.}\/'s distillation process~\cite[Fig.14]{bennett96m} 
using a quantum code with no encoding map
can be visualized as
in %the quantum-circuit-like diagram of 
Figure~\ref{fig:1}.
%
%figure 1 here?
%
Suppose we are given a quantum code $(\Cq \subset \Hch^{\tnsr n}, \dec)$
with $\dim \Cq =K$, which works on the teleportation channel
$\tch{\shs_n}$,
a $K$-dimensional Hilbert space
$\Hch_{n, {\rm R}}$
and $n$ copies of $\Ht \simeq \Hch$ on site A,
and a purification of $\hPi_{\Cq}$, i.e.,
a maximally entangled state
$
\ket{\men{\cB_{\rm R},\cB_{\rm T}}}
$
where $\cB_{\rm R}=\{ \ket{j}_{\rm R} \}_{0\le j< K}  \subset \Hch_{n,{\rm R}}$ and
$\cB_{\rm T} = \{ \ket{j} \}_{0\le j< K} \subset \Cq \subset \Ht^{\tnsr n}$
are orthonormal systems.
Here we identify
$\Htn$ and $\Hbn$ with $\Hch^{\tnsr n}$ via isomorphisms.
%Put $\Haux{n,{\rm A}}=\Hch_{n,{\rm R}}\tnsr\Ht^{\tnsr n}$
We perform the teleportation protocol in Section~\ref{ss:telechan} with
the bipartite state $\shs_n\in\Bop(\Han\tnsr\Hbn)$ leaving 
the system of $\Hch_{n,{\rm R}}$ untouched, and
then the recovery operation $\dec$ on the site B.
The role of the code $(\Cq \subset \Hch^{\tnsr n}, \dec)$ is to transmit 
entanglement shared with the system of $\Hch_{n,{\rm R}}$
from $\Htn$ to $\Hbn$, and we see
by the definition of entanglement fidelity,
the fidelity of this distillation protocol is given by
that of the code $(\Cq, \dec)$:
\begin{equation}\label{eq:fid_dc}
\bra{\men{\cB_{\rm R},\cB_{\rm B}}} \Dis_n(\shs_n) \ket{\men{\cB_{\rm
R},\cB_{\rm B}}} = \Fen( \hPi_{\Cq}, \dec\circ\tch{\shs_n}),
\end{equation}
where $\Dis_n(\shs_n)= \trace_{\sK} [(\Id_{\rm RTA} \tnsr \dec) \circ (\Id_{\rm R} \tnsr \Tel)]
(\ket{\men{\cB_{\rm R},\cB_{\rm T}}} \bra{\men{\cB_{\rm R},\cB_{\rm T}}} \tnsr
\shs_n ) $, $\sK={\Htn\tnsr\Han}$,
and $\cB_{\rm B}$ is the image of $\cB_{\rm T}$ under the isomorphism.
We summarize the above argument in the following statement.
\begin{theorem}\label{th:main}
Let $\Hch$ be a finite-dimensional Hilbert space. %CANDIDATE4OMIT?
Whenever a quantum code 
$(\Cq \subset \Hch^{\tnsr n}, \dec: \Bop(\Hch^{\tnsr n})\to
\Bop(\Hch^{\tnsr n}))$ with $K =\dim \Cq$ exists,
we have
\[ %\label{eq:main}
\Fdis{K}{\shs_n} \ge  \Fen( \hPi_{\Cq}, \dec\circ\tch{\shs_n})
\]
for any state $\shs_n\in \Bop({\Hch^{\tnsr n}}\tnsr{\Hch^{\tnsr n}})$.
In other words,
\[
\Fdis{K}{\shs_n} \ge \Fch{n}{K}{\tch{\shs_n}}
\]
for any $0< K \le \dmn^n$.
\end{theorem}

This implies for any $\{\shs_n\}$,
\begin{equation}\label{eq:QleD}
\Capa(\{\tch{\shs_n}\}) \le \Dopt(\{\shs_n\}),
\end{equation}
which was known for $\shs_n = \shs^{\tnsr n}$~\cite{bennett96m}.
Clearly, Theorem~\ref{th:main} as well as (\ref{eq:QleD}) is true for
an arbitrary teleportation scheme~\cite{werner01}
or any other operation that defines a map $\tch{\shs_n}$ similarly
and is allowed in the presupposed class $C$ of distillation protocols,
though concrete expressions for $\tch{\shs_n}$ seem unknown except
for the teleportation with $\{ \Ebe_y \}_{y\in\Zdpower{\dmn}{2n}}$.

The bounds~\cite{hamada01e,hamada02m,hamada02c}
on $1-\Fen(\hPi_{\Cq}, \dec\circ\cA_n)$
to be used in the next section were originally claimed
in terms of minimum (pure-state) fidelity in place of 
entanglement fidelity $\Fen$. However, we have the following lemma.
\begin{lemma} \label{lem:fidelities} \cite[Theorem~2]{barnum00}
For any TPCP linear map $\cM$, 
\[
1-\min_{\ket{\phi}\in\Cq} 
\bra{\phi}\cM(\ket{\phi}\bra{\phi})\ket{\phi} \le G,
\]
where $ \ket{\phi}$ is normalized, implies
\[
1-\Fen(\hPi_{\Cq},\cM) \le \frac{3}{2} G.
\]
\end{lemma}

In most asymptotic settings, 
the factor $\frac{3}{2}$ is negligible.
Using Lemma~\ref{lem:telechan} and \ref{lem:fidelities}, we obtain 
the following corollary to Theorem~\ref{th:main} %or (\ref{eq:main}),
where $\Ebe_J=\{ \Ebe_x \mid x\in J \}$ for $J\subset \cX^n$
and the term `$\Ebe_J$-correcting' is in the sense of Knill and Laflamme~\cite{KnillLaflamme97}.
\begin{corollary}\label{coro:symp4dis}
Let $(\Cq,\dec)$ be an $\Ebe_J$-correcting code 
with $K =\dim \Cq$.
Then,
\[
1-\Fdis{K}{\shs_n} \le \frac{3}{2} \sum_{x\in\SINT_{\dmn}^{2n}:\, x\notin J} P_n(x)
\]
for a bipartite state
$\shs_n\in \Bop({\Hch^{\tnsr n}}\tnsr{\Hch^{\tnsr n}})$,
where $P_n(x)$
is given by
\[
P_n(x)=\bra{\Psi_{x}} \shs_n \ket{\Psi_{x}},
\quad x\in\Zdpower{\dmn}{2n}.
\]
\end{corollary}

Note that this corollary allows any correlation in $\shs_n$ among
the $2n$ factor systems.
Note also that widely investigated symplectic codes~\cite{Gottesman96,crss97}
%,crss98} 
have enough flexibility
to cope with such general states $\shs_n$ in principle.
To see this, recall that a symplectic code is obtained from
a subspace $\Cso \subset \Zdpower{\dmn}{2n}$ that are contained in
the symplectic dual $\Cso^{\perp}$ of $\Cso$.
If we choose a vector $\hat{x}(s)$ from each coset $s$ of $\Cso^{\perp}$
in $\Zdpower{\dmn}{2n}$,
and denote the set of coset representatives $\hat{x}(s)$ by $J_0$,
we have $\Ebe_{J}$-correcting symplectic codes $(\Cq,\dec)$
where $J=J_0+\Cso$
(see \cite{hamada02c} for a self-contained exposition).
A natural choice for $\{ \hat{x}(s) \}$ is
one that, at least nearly, maximizes
$P_n(J) = \sum_{x\in J} P_n(x)$.
In fact, the bounds on $\Fch{n}{\dmn^{Rn}}{\cA_n}$ 
in Section~\ref{ss:channels} below were proved with
such choices. %of $\hat{x}(s)$.
Observe that nothing was assumed here on the probability distribution $P_n$
or on the state $\shs_n$ to be distilled.
It is remarked that if $\Cq$ is a symplectic code, the factor $\frac{3}{2}$
in the bound in Corollary~\ref{coro:symp4dis} can be removed
since $\Fen( \hPi_{\Cq}, \dec\circ\tch{\shs_n})=P_n(J)$ for the above set $J$
and the corresponding codes $(\Cq,\dec)$.

\section{Known Results on Quantum Codes and Their Implications \label{ss:channels}}

In this section, 
we give more concrete exponential lower
bounds on the fidelity of distillation
and those on $\Dopt(\{ \shs_n \})$ %as their consequences
assuming $\dmn$ is a prime. 

\subsection{Distillation of Independent Mixed States}

We begin with the easy case in which $\shs_n$ are the tensor power
$\shs^{\tnsr n}$ of a state $\shs\in \Bop(\Ha\tnsr\Hb)$.
By Lemma~\ref{lem:telechan},
the teleportation channel $\tch{\shs_n}$ can be written as
$\tch{\shs}^{\tnsr n}$ with $\tch{\shs} \sim \{ \sqrt{P(u)}
\Ebe_{u}\}_{u\in\cX}$, where $P=P_\shs$ is the probability
distribution on $\cX$ given by
\[
P_\shs(u)=\bra{\Psi_{u}} \shs \ket{\Psi_{u}}, \quad u\in\cX=\Zdpower{\dmn}{2}.
\]

A known exponential bound for the channel $\{ \tch{\shs}^{\tnsr n} \}$
has the form 
\begin{equation} \label{eq:exp_gen2}
 1 - \Fch{\varM}{\dmn^{R\varM}}{\tch{\shs}^{\tnsr\varM}} 
\le \exp_{\dmn}[
- n \sup_{\Cso} {\hEnk{\varnin}{\varkin}{\Cso}(R,P_{\shs})} + o(n) ],
\end{equation}
where $\exp_{\dmn}x=\dmn^{x}$ and 
the supremum is taken over all self-orthogonal even-length
subspaces $\Cso$ 
of $\Zdpower{\dmn}{2n}$ with $\dmn$ prime~\cite{hamada02c}.
We will not give the
specifications of
$\hEnk{\varnin}{\varkin}{\Cso}(R,P)$ here
but only recall that
when $\Cso$ consists solely of the zero vector, 
%in which case the inner code does nothing,
$\hEnk{\varnin}{\varkin}{\Cso}(R,P)$ is the same as the simple one
$E(R,P)=\min_{Q} [ D(Q||P) + |1-H(Q)-R|^+ ]$
(see \cite{hamada01e,hamada01g} for the details),
which still gives the best numerical lower bound known for most channels.
%~\cite{hamada01e,hamada01g}
%(graphs for some channels are given in Figs.~1 of them).
\begin{comment}
\[
E(R,P)=\min_{Q} [ D(Q||P) + |1-H(Q)-R|^+ ],
\]
$|x|^+ =\max\{x,0\}$, $H$ and $D$ denote classical entropy and 
relative entropy, respectively, and
the minimum is taken over all probability distributions on $\cX$. 
\end{comment}

These bounds are
applicable to entanglement distillation through Theorem~\ref{th:main}:
%For any $0\le R \le 1$, %4SPACE
\begin{eqnarray}
1-\Fdis{\dmn^{nR}}{\shs^{\tnsr n}} &\le& \exp_{\dmn}[- n \sup_{\Cso} \hEnk{\varnin}{\varkin}{\Cso}(R,P_\shs) + o(n) ] \label{eq:C1}\\
&\le&\exp_{\dmn} [ -n E(R,P_\shs)+ o(n) ] \label{eq:C2}.
\end{eqnarray}
The lower bounds on $\Dopt(\{ \shs^{\tnsr n} \})$ that directly follow from
these %exponential %4SPACE
bounds are
\begin{eqnarray}
\Dopt(\{ \shs^{\tnsr n} \}) &\ge & \sup_{m>0} \, \sup_{\Cq:\,\, {\rm symplectic\,\, code}\,\,\subset \Hch^{\tnsr m}}
I_{\rm c}(\hPi_{\Cq}, \tch{\shs}^{\tnsr m})/m \label{eq:D1} \label{eq:lbIc}\\
&\ge & I_{\rm c}(\hPi_{\Hch}, \tch{\shs}) = 1-H(P_{\shs}), \label{eq:D2}
\end{eqnarray}
where $I_{\rm c}$ and $H$ denote coherent information and entropy, 
respectively, the bound in (\ref{eq:D2}) follows from (\ref{eq:C2})
and that $E(R,P_\shs)>0$ whenever $R<1-H(P_{\shs})$~\cite{hamada01e},
and (\ref{eq:D1}) follows from (\ref{eq:C1}) similarly~\cite{hamada02c}.
In (\ref{eq:lbIc}), the supremum is taken over all symplectic codes
designed with the same basis $\{ \Ebe_x \}_{x\in\cX}$ as we 
use for the teleportation protocol.

\subsection{Distillation of Mixed States with Classical Markovian Correlation \label{ss:disMC}}

In this subsection,
we consider a sequence of states $\{\shs_n\}$ with 
certain correlation.
%classical correlation of Markovian nature.?
For a motivation, imagine, by means of
a Gedankenexperiment, we were given a sequence of states $\{ \shs_n \}$
and could perform the bipartite measurement
\[ %\begin{eqnarray}
\{ \ket{\Psi_{x}}\bra{\Psi_{x}} \}_{x\in\Zdpower{\dmn}{2n}},
\] %\end{eqnarray}
i.e., $n$ trials of the bipartite measurement
$
\{ \ket{\Psi_{u}}\bra{\Psi_{u}} \}_{u\in\Zdpower{\dmn}{2}}
$.
Then, we would obtain, as measurement results, random variables 
$\sX_1,\dots,\sX_n$ 
taking values in $\Zdpower{\dmn}{2}$
such that $\Prob \{ (\sX_1,\dots,\sX_n)= x \}
= \bra{\Psi_x} \shs_n \ket{\Psi_x} = P_n(x)$.
This distribution is exactly the same as that appearing in the expression
for $\tch{\shs_n}$ in Lemma~\ref{lem:telechan}.
Those $\{ \shs_n \}$ for which $\sX_1,\dots,\sX_n$ are independent
identically distributed random variables, viz.,
$P_n(x_1,\dots,x_2)=P(x_1)\dots P(x_n)$,
have been treated in the previous subsection.
We will consider a more general case, where
$\sX_1,\sX_2,\dots,$ form a homogeneous
(first-order) Markov chain, viz.,
\begin{equation}\label{eq:mcP}
 P_n(x_1,\dots, x_n)= p(x_1) \prod_{j=1}^{n-1} P(x_{j+1}| x_{j}) \label{eq:Mar}
\end{equation}
for some transition probabilities $P(v|u)$, $u,v\in\cX$,
and some initial distribution $p$.
If the Markov chain is irreducible, then 
by Theorem~\ref{th:main} and the known bound
on $\Capa(\{ \tch{\shs_n} \})$~\cite{hamada02m},
we have
\begin{equation}\label{eq:mc}
\Dopt(\{ \shs_n \}) \ge 1-H(P|q)
\end{equation}
where $H(P|q)$ is the entropy of $P(\cdot|\cdot)$ conditional on
the unique stationary distribution $q$ of the Markov chain.
For an attainable error exponent, see \cite{hamada02m}.

{\em Example 1.}\/
Let us assume $\dmn=2$
and define $P(v|u)$ by 
\[
\mbox{$P((0,0)|u)= 1- \vep_u$}\quad \mbox{and} \quad \mbox{$P(v|u)=\vep_u/3$\, for\, $v\ne(0,0)$},
\]
$0<\vep_u<1$, $u\in\cX$. 
Then, (\ref{eq:mc}) becomes
\[
\Dopt(\{\shs_n\}) \ge 1- 
\sum_{u\in\cX} q(u) [ h(\vep_u) + \vep_u \log_2 3 ],
\]
where $h(z)=-z \log_2 z -(1-z) \log_2 (1-z)$.
Note that when $\vep_u$ is independent of $u$, 
$\{ \shs_n \}$ is a sequence of
identical isotropic states, and the %lower %4SPACE
bound becomes
the known one~\cite{bennett96m}. 
\enproof

The bound in (\ref{eq:mc}) is also true when
the support of the initial distribution $p$ is  
contained in an equivalence class $\cX'\subset\cX$,
where the equivalence relation holds between $u$ and $v \in \cX$ 
if $u$ and $v$ lead to each other,
i.e., if $P^{(n)}(v|u)>0$ and $P^{(m)}(u|v)>0$
for some positive integers $n$ and $m$ 
with $[ P^{(n)}(v|u) ]$ being the $n$-th power of 
the matrix $[ P(v|u)]$~\cite{chung};
in this case, $q$ is to be understood as
the unique stationary distribution $q$ whose support is $\cX'$.
To see this, it is enough to notice that the probability of
occurrence of any $x\notin \cX'$ vanishes, so that we can safely
replace $\cX$ by $\cX'$ in deriving the bound~\cite{hamada02m}.

{\em Example 2.}\/
Let us assume
$\shs_n$ can be written in the form 
\[ %begin{equation}\label{eq:diagonalchan}
\shs_n = \sum_{x\in\cX^n} P_n(x) \ket{\Psi_{x}}\bra{\Psi_{x}}
\] %end{equation}
with (\ref{eq:mcP}), 
there exists an equivalence class $\cX'$
which is contained in $\{ (0,v) \mid v\in\SINT_\dmn \}$,
and the initial distribution $p$ is the stationary one $q$ 
whose support is $\cX'$.
(The corresponding channel $\tch{\shs_n}$ is
a probabilistic mixture of conjugation actions of 
tensor products of several $Z$ and $I$.)
Then, the bound (\ref{eq:mc}) is tight for this sequence of states
(even in the wider class of PPT distillation protocols),
which has been known for the special case where the Markov chain
is a sequence of independent, identically distributed random variables~\cite{rains97,rains01,VollbrechtW02}.
This can be shown by examining the derivation of
Rains' upper bound~\cite{rains98}.
As a byproduct, 
$\Capa(\{ \tch{\shs_n} \})=1-H(P|q)$ is concluded by (\ref{eq:QleD})
for this example.
\enproof

\section{Concluding Remarks \label{ss:conc}}

We have presented an explicit formula of the teleportation channel
$\tch{\shs_n}$
using Weyl's projective representation of $(\SINT/\dmn\SINT)^{2n}$,
and have seen that
the optimum fidelity $\Fdis{K}{\shs_n}$ of a one-way distillation scheme
is lower-bounded by 
that of quantum codes $\Fch{n}{K}{\tch{\shs_n}}$.
An application of these results has shown that
reliable distillation from correlated mixed states is possible.

Symplectic codes can be used for entanglement distillation
in another way, as argued by Shor and Preskill in a security proof
for quantum key distribution~\cite{ShorPreskill00}. %4SPACE
%of a quantum key distribution protocol~\cite{ShorPreskill00}.
Using this scheme and (\ref{eq:twirling}),
we can reproduce the lower bounds on
$\Fdis{K}{\shs_n}$ in Section~\ref{ss:channels},
though the derivation applies only to a restricted class of codes 
%symplectic codes (or possibly to a larger algebraic class) 
and hence does not imply Theorem~\ref{th:main} or (\ref{eq:QleD}).

The problem of determining the optimum achievable rates 
is not yet settled 
even for distillation from identical copies of bipartite states, 
whether the allowed protocols are one-way or two-way
(or seemingly more tractable PPT ones), and further investigation is awaited
(e.g.,~\cite{rains01,HHH00u}).
Recently, P.\ Shor announced
that the known upper bound on the quantum capacity
written with coherent information~\cite{barnum00}
is actually the quantum capacity 
(MSRI Workshop, MSRI, Berkeley, California, Nov.\ 2002).
This implies
$\Dopt(\{ \shs^{\tnsr n} \})
\ge $\\ $ %4LAYOUT
\sup_{m>0} \sup_{\DOgen \in \Bop(\Hch^{\tnsr m}):\,\, {\rm state}}$
$I_{\rm c}(\DOgen, \tch{\shs}^{\tnsr m})/m = 
\sup_{m>0} \sup_{\Cq:\,\, {\rm subspace\,\, of}\,\, \Hch^{\tnsr m}}
I_{\rm c}(\hPi_{\Cq}, \tch{\shs}^{\tnsr m})/m$,\\ %4LAYOUT%QIC
where the equality is due to Lemma~1 of \cite{bsst02e} (also \cite{holevo02e}).
It is not known if this bound is strictly 
greater than that in (\ref{eq:lbIc}).

\section*{Acknowledgments}

The author wishes to thank D.~W.~Leung for a useful conversation
on a discrete twirling.
He is grateful to K.~Matsumoto and H.~Imai of the QCI project for support.

\section*{References}

%{\small 

%}

\end{document}